\documentclass[twoside,leqno,twocolumn]{article}

\usepackage[letterpaper]{geometry}
\usepackage{ltexpprt}
\usepackage{graphicx}
\usepackage[utf8]{inputenc}
\usepackage[english]{babel}
\usepackage{amsmath}
\usepackage{listings}
\usepackage{hyperref}
\usepackage{bbding}
\usepackage{pifont}
\usepackage{wasysym}
\usepackage{amssymb}
\usepackage{tabularx}

\lstdefinestyle{mystyle}{
    language=Python, 
    frame = single, 
    basicstyle=\ttfamily\footnotesize,
    breakatwhitespace=true,         
    breaklines=true,                 
    captionpos=b,                    
    keepspaces=true,                 
    numbers=left,                    
    numbersep=5pt,                  
    showspaces=false,                
    showstringspaces=false,
    showtabs=false,                  
    tabsize=2
}

\begin{document}

\title{\Large Context-Aware Drive-thru Recommendation Service at Fast Food Restaurants}
\author{Luyang Wang\thanks{Burger King Corporation. (lwang1@whopper.com)}
\and Kai Huang\thanks{Intel Corporation. (\{kai.k.huang,  jiao.wang, shengsheng.huang, jason.dai\}@intel.com)}
\and Jiao Wang\footnotemark[2]
\and Shengsheng Huang\footnotemark[2]
\and Jason(Jinquan) Dai\footnotemark[2]
\and Yue Zhuang\thanks{Brown University. (yue\_zhuang1@brown.edu)}}

\date{}

\maketitle







\begin{abstract} \small\baselineskip=9pt
Drive-thru is a popular sales channel in the fast food industry where consumers can make food purchases without leaving their cars. Drive-thru recommendation systems allow restaurants to display food recommendations on the digital menu board as guests are making their orders. Popular recommendation models in eCommerce scenarios rely on user attributes (such as user profiles or purchase history) to generate recommendations, while such information is hard to obtain in the drive-thru use case. Thus, in this paper, we propose a new recommendation model \emph{Transformer Cross Transformer} (T\emph{x}T), which exploits the guest order behavior and contextual features (such as location, time, and weather) using Transformer encoders for drive-thru recommendations. Empirical results show that our T\emph{x}T model achieves superior results in Burger King's drive-thru production environment compared with existing recommendation solutions. In addition, we implement a unified system to run end-to-end big data analytics and deep learning workloads on the same cluster. We find that in practice, maintaining a single big data cluster for the entire pipeline is more efficient and cost-saving. Our recommendation system is not only beneficial for drive-thru scenarios, and it can also be generalized to other customer interaction channels.
\end{abstract}

\section{Introduction}
Modern restaurant drive-thru service is a type of take-out service that allows customers to purchase food without leaving their cars. The process starts where customers browse the items on the outdoor digital menu board and talk to the cashier inside the restaurant to place their order. Drive-thru recommendation systems have recently drawn the attention of large players in the quick serving restaurant industry. At Burger King, we take the initiative to develop a state-of-the-art context-aware recommendation system to improve our guest experience in the drive-thru.

Drive-thru being an offline shopping channel has much fewer data points that can be leveraged to generate proper recommendations. For example, there is no easy way to reliably identify guests and retrieve their profiles in the drive-thru environment. Lack of user identifier makes popular recommendation algorithms such as Alternating Least Squares (ALS) described in \cite{koren2009matrix} and Neural Collaborative Filtering (NCF) \cite{he2017neural} not applicable for drive-thru as they require user profiles as model inputs. Session-based recommendation systems are able to learn from behavior sequence data to generate recommendations so they are less dependent on user identifier and its related information. However in our drive-thru use case, only relying on order sequence data for prediction is simply not enough. Fast food purchase preference can drastically change given different context data like location, time, and current weather conditions. A proper neural network architecture that leverages both order sequence data and multiple context features is key to a successful recommendation system in the drive-thru scenario.

To tackle aforementioned challenges, we propose a new deep learning recommendation model called \emph{Transformer Cross Transformer} (T\emph{x}T) that exploits the sequence of each order as well as the context information to infer a customer's preference at the moment. Sequential signals underlying the guest’s current order are important to reflect behavioral patterns. Using apple pie as an example, an apple pie shown earlier during an order sequence suggests this is a lightweight snack-driven order while an apple pie ordered after multiple large sandwiches could imply this is a large multi-party order. Contextual data such as time and weather could also be useful since guests may have particular preferences according to the current situation. For example, it is natural that people tend to have cold drinks when the temperature increases. Furthermore, we incorporate other context information such as the store location so that our recommendation can prioritize popular food items being sold in particular regions.

In addition to designing a suitable algorithm for our drive-thru scenario, setting up a complete pipeline to efficiently train the model on our enormous guest transaction data is also important for the whole system. We do have big data platforms to store and process the big data, but how to seamlessly integrate deep learning training to the existing big data system is a nontrivial task. A common practice is to allocate a separate system dedicated for distributed training only, which introduces extra data transfer time and system maintenance efforts. Instead, we propose an integrated system allowing big data analytics and distributed training on exactly the same cluster where the data is stored, which is efficient, easy to scale and maintain.

In summary, the main contributions of this paper are as follows:

\begin{itemize}
\setlength\itemsep{0.5em}
\item We propose a \emph{Transformer Cross Transformer} model (T\emph{x}T) for the drive-thru scenario. The key advantage of our model is that we apply Transformer encoder \cite{vaswani2017attention} to capture both user order behavior sequence and complicated context features and combine both transformers through latent cross \cite{beutel2018latent} to generate recommendations. 

\item We design a unified system architecture based on Apache Spark \cite{zaharia2012resilient} and Ray \cite{moritz2018ray} to perform distributed deep learning training on exactly the same cluster where our big data is stored and processed. Our system fully utilizes the existing big data clusters, avoids the expensive data transfer, eliminates separate workflows and systems, and thus improves the end-to-end efficiency in the production environment.

\item We have successfully deployed our recommendation system at Burger King’s restaurants to serve our drive-thru customers. We share our key learnings in production and have open sourced our implementation in \cite{AnalyticsZoo}. Our model concepts and system design could be easily extended to other customer interaction channels.
\end{itemize}

\section{Related Work}
In this section, we give a brief overview of the past research work in terms of recommendation that we gain insights from. Our work extends state-of-the-art models of session-based recommendation and context-aware recommendation to make the topology best fit our drive-thru scenario.

\subsection{Session-based Recommendation.}

A straightforward approach in session-based recommendation is item-based collaborative filtering \cite{sarwar2001item}, which uses an item-to-item similarity matrix to recommend the most similar items to the one that the user has clicked most recently in the session. Another approach is mainly based on Markov chains (e.g. Markov Decision Processes \cite{shani2005mdp}) which capture user behavior sequential patterns to predict a user’s next action given the previous actions. In recent years, RNN based neural networks including Long Short-Term Memory (LSTM) \cite{hochreiter1997long} and Gated Recurrent Unit (GRU) \cite{cho2014learning} have been adopted and yield better recommendation results. For example, GRU4Rec \cite{Hidasi2016SessionbasedRW} utilizes session-parallel mini-batch training and ranking-based loss function for recommendation. Later, other RNN based algorithms including Improved GRU4Rec \cite{Hidasi2018RecurrentNN} and user-based GRU \cite{donkers2017sequential} are explored for session-based recommendation as well.

Attention mechanism has become another powerful approach to modeling sequential signals in a user session. There are studies leveraging attention to session-based recommendation. For example, NARM \cite{li2017neural} hybrids attention into GRU to capture the behavior and purpose of users in a session. Instead of incorporating attention as a component into other models, Transformer \cite{vaswani2017attention} networks, which only use multi-head self-attention to model sequential data, achieve state-of-art performance (e.g. SASRec \cite{kang2018self}). BST \cite{chen2019behavior} uses a Transformer layer to learn item representations in a behavior sequence and concatenates the result with other feature embeddings for the final prediction.

\subsection{Context-Aware Recommendation.}

Traditional recommendation systems deal with applications which only include users and items. But in many scenarios, only user and item attributes may not be sufficient, and it is also important to take contextual information into consideration to recommend items to users under certain circumstances. This is more important for fast food recommendation since food order habits of users may change significantly given current location, time and weather conditions.

More recently, many works have started on context-aware recommender systems (CARS) \cite{adomavicius2011context} to incorporate available context features into the recommendation process as explicit additional categories of data. In Wide \& Deep Learning \cite{he2017neural}, context features can be treated as categorical features and fed into hidden layers of a linear model. \cite{twardowski2016modelling} applies general context features to the RNN layer. Most commonly used approaches concatenate contextual features to other inputs. In the meantime, Latent Cross \cite{beutel2018latent} performs an element-wise product between the context embedding and the hidden states from RNN to improve the performance of context incorporation. But the paper doesn’t focus on how to effectively learn the feature representations from multiple complex context features. A major difference of our proposed model in this paper is that we apply a separate Transformer block to capture the complicated interactions of different context features.

\section{Model Description}
In this section, we describe our \emph{Transformer Cross Transformer} model (T\emph{x}T) in detail. We begin with defining a drive-thru recommendation task followed by illustrating the structure of our T\emph{x}T model.

\subsection{Problem Description.}

We define the drive-thru recommendation task as follows. Given an order sequence event $S_n(e)=\{f_1,f_2, ...,f_n\}$ where $f_i$  is the $i$-th food item a guest orders, and context feature collection $C(e)=\{c_1,c_2, ...,c_m\}$ where $c_i$ is one context feature when the order event occurs (e.g. location, time and weather), we shall learn a function, $F$,

\vspace{-1mm}
\begin{equation*}
f_{n+1} = F(S_n(e), C(e))
\end{equation*}

which predicts the food item the guest is most likely to order next, i.e. $f_{n+1}$.

\begin{figure*}[ht]
\centering
\includegraphics[width=0.9\linewidth]{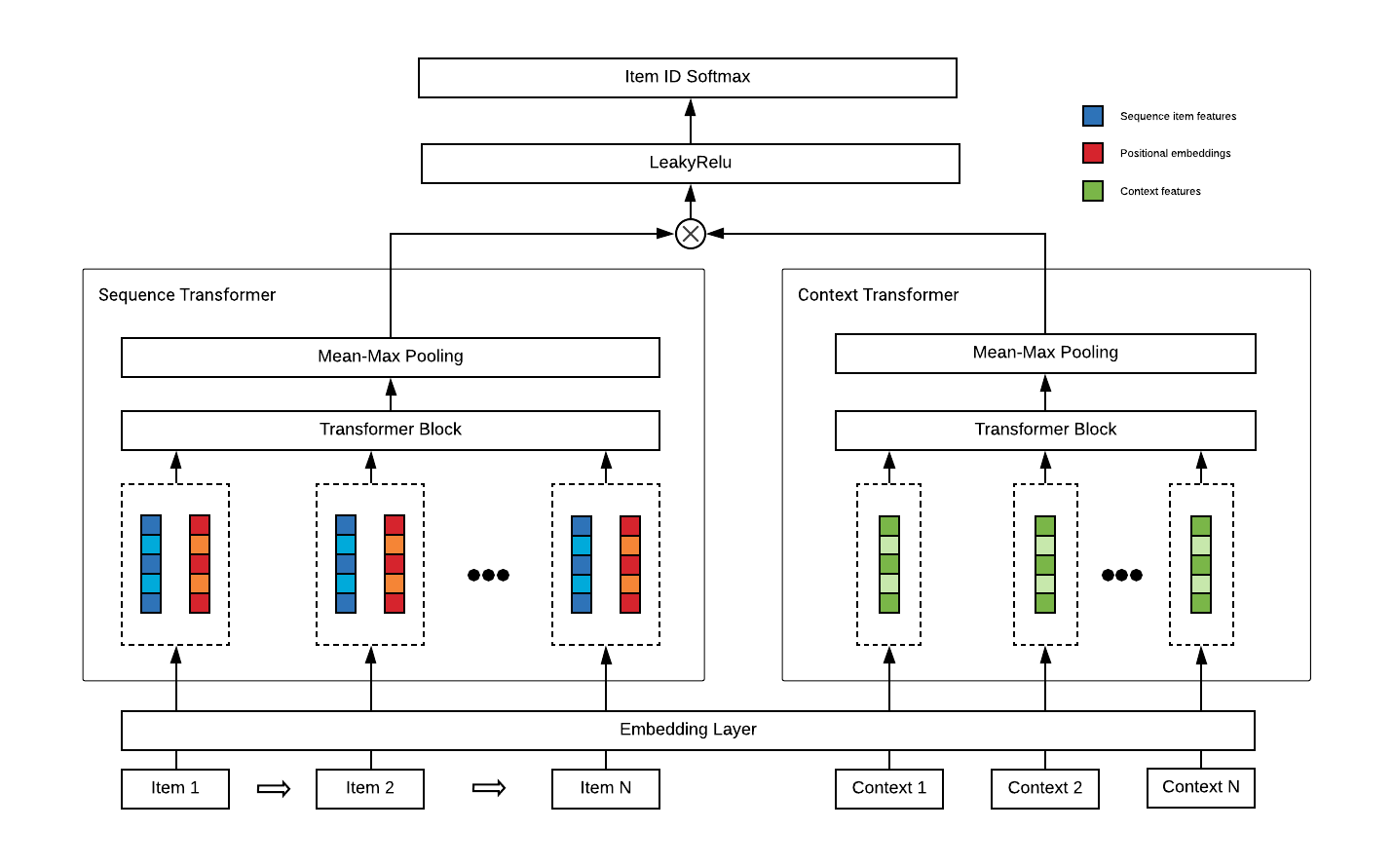}
\caption{The model structure of T\emph{x}T.}
\label{fig:model}
\end{figure*}

To effectively deal with the drive-thru recommendation task, we propose the \emph{Transformer Cross Transformer} model (T\emph{x}T), which uses a \emph{Sequence Transformer} to encode guest order behavior, a \emph{Context Transformer} to encode context features,  and then uses an element-wise product to combine them to produce the final output, as shown in Figure \ref{fig:model}.

\subsection{Sequence Transformer.} \label{sequencetransformer}

We construct a \emph{Sequence Transformer} to learn the sequence embedding vector of each item in the guest order basket, as shown in the lower left part of Figure \ref{fig:model}. To ensure that the item position information can be considered in its original add-to-cart sequence, we perform positional embedding on input items in addition to the item feature embedding. Both embedding outputs are added together and fed into a multi-head self-attention network \cite{vaswani2017attention}.

In the multi-head self-attention network, we project item embedding into a set of query $\mathbf{Q}$ and key-value pairs ($\mathbf{K}$; $\mathbf{V}$) to $h$ spaces; and then calculate the multi-head self-attention output in parallel:

\vspace{-3mm}
\begin{equation*}
\begin{gathered}
MultiHead\left(\mathbf{Q}, \mathbf{K}, \mathbf{V}\right) = Concat\left(head_1,...,head_h\right)\mathbf{W}^O \\
head_i = Attention\left(\mathbf{Q}, \mathbf{K}, \mathbf{V}\right) = softmax\left(\frac{\mathbf{Q}\mathbf{K}^\intercal}{\sqrt{d_k}}\right) \mathbf{V}
\end{gathered}
\end{equation*}

where $d_k$ is the dimension of $\mathbf{K}$, $h$ is the number of heads and $\mathbf{W}^O$ is the weight of the heads.

Padding mask is applied where attention scores at padded item positions are replaced with infinitive negative values so that after the softmax, the attention weights at these padded positions will be close to 0. We then add point-wise Feed-Forward Networks on top of the multi-head self-attention outputs. 

To extract the vector representation of the whole guest order basket information from the hidden vectors of each item, we need to apply a pooling function here. Mean pooling is a common pooling strategy which performs element-wise mean calculation across all product vectors such that the output basket embedding is represented by all the products contained in the product sequence. Max pooling is another popular pooling strategy which picks the maximum value at each position across all product vectors and therefore the output basket embedding is represented by a small number of key products and their salient features only. To combine the advantage of both mean pooling and max pooling, we follow mean-max pooling \cite{zhang2018learning} to apply mean pooling and max pooling separately against the output of the Transformer block and concatenate both pooling outputs to form the guest order basket vector. The final output of the \emph{Sequence Transformer} can be formalized as:

\begin{equation*}
\begin{gathered}
\alpha =sum(z_i) /n; \beta = max(z_i) \\
output = Concat(\alpha, \beta)
\end{gathered}
\end{equation*}

where $\mathbf{Z}=\{z_i\}$ is the output from the Transformer encoder layer.

\subsection{Context Transformer.} \label{contexttransformer}

Context information is a significant factor in drive-thru recommendation scenarios. A common way to incorporate contextual features is to directly concatenate context features with other inputs. But it is less meaningful simply concatenating non-sequence features with sequence features. In \cite{beutel2018latent}, the author proposed an element-wise sum when dealing with multiple context features before crossing to the sequence output. However, sum can only represent how context features aggregately contribute to the output, but can not represent the individual contribution of each context feature. For example, the restaurant location may sometimes have a stronger impact on guest food ordering compared with other context features due to regional food offerings. In this case, a simple sum of all the context features cannot emphasize the importance of location. 

\begin{figure}[ht]
\centering
\includegraphics[width=0.9\linewidth]{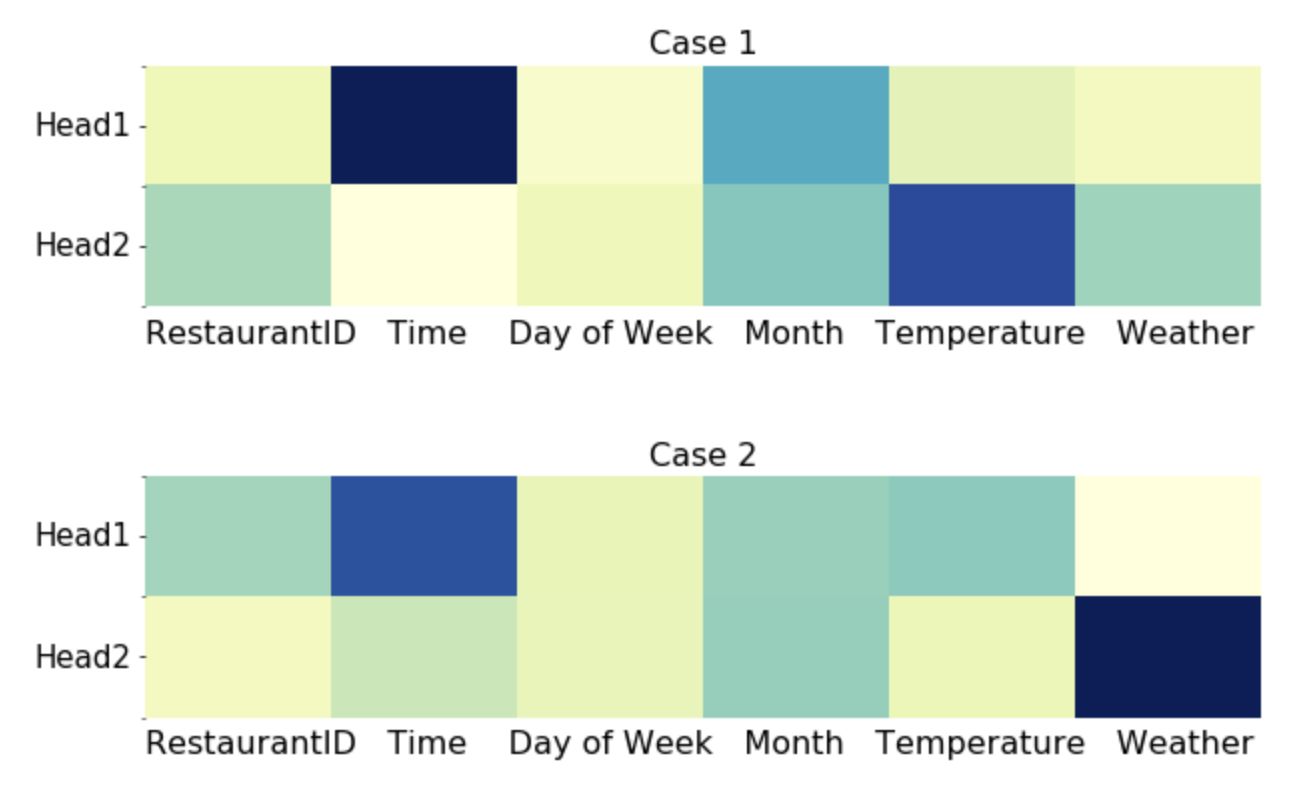}
\caption{Heat map for context features}
\label{fig:visual}
\end{figure}

Therefore, we use a \emph{Context Transformer} to encode the contextual information, as shown in the bottom right part of Figure \ref{fig:model}. Using Transformer’s multi-head self-attention \cite{vaswani2017attention}, we can capture not only the individual effect of each context, but also the internal relationship and interactions across different context features. Attention outputs of different heads focus on the importance of different context features, which enables our model to consider the combined influence of various context features at the same time. In Figure \ref{fig:visual}, the color depth of boxes indicates the weights of context features in terms of each head and the darker ones are treated as more important features. From Case 1 in Figure \ref{fig:visual} we can see that Head 2 focuses on Temperature while Head 2 in another Case 2 focuses more on Weather.

In particular, we assemble multiple context embeddings into a context vector and apply multi-head self-attention to it. Unlike the \emph{Sequence Transformer}, we don’t create positional-encoding to the input of multi-head self-attention since the order of context features in the context vector is fixed in our model. In production we find that one Transformer layer is sufficient to well extract our context features. Similar to \emph{Sequence Transformer}, we also apply mean-max pooling \cite{zhang2018learning} on the Transformer output to generate the overall representation of contextual features.

\subsection{Transformer Cross Transformer.}

To jointly train \emph{Sequence Transformer} and \emph{Context Transformer}, we adopt the idea proposed in \cite{beutel2018latent} to perform an element-wise product between these two outputs. Through this cross Transformer training, we are able to optimize all the parameters such as item embeddings, contextual feature embeddings and their interactions at the same time. Finally, we apply LeakyRelu as the activation function followed by a softmax layer to predict the probabilities of each candidate item.

\section{Distributed Deep Learning Architecture}

\subsection{Motivations.}
We have a huge amount of  user transaction records stored on our big data cluster, which is impossible to train on a single node. So we aim to build an end-to-end training pipeline containing a distributed data cleaning and ETL process, which we implement using Apache Spark \cite{zaharia2012resilient}, and a distributed training process to train our T\emph{x}T model.

Conventional approaches to build such a pipeline would normally set up two separate clusters, one dedicated to big data processing, and the other dedicated to deep learning training (e.g., a GPU cluster). Unfortunately, this not only introduces a lot of overhead for data transfer, but also requires extra efforts for managing separate workflows and systems in production. In addition, while popular deep learning frameworks \cite{abadi2016tensorflow, paszke2017automatic, chen2015mxnet} and Horovod \cite{sergeev2018horovod} from Uber provide support for data parallel distributed training (using either parameter server architecture \cite{li2014scaling} or MPI \cite{graham2006open} based AllReduce), it can be very complex to properly set them up in production. For instance, these methods usually require all relevant Python packages pre-installed on every node, and the master node has SSH permission to all the other nodes, which is probably infeasible for the production environment and also inconvenient for cluster management. 

To address these challenges, we propose and implement a unified system (i.e. \emph{RayOnSpark}) using Spark and Ray \cite{moritz2018ray}, which runs the end-to-end data processing and deep learning training pipeline on the same big data cluster. This unified system architecture has greatly improved the end-to-end efficiency of our production applications, and is open sourced as a part of the project \cite{AnalyticsZoo}.

\subsection{RayOnSpark.} \label{rayonspark}

\begin{figure}[ht]
\centering
\includegraphics[width=0.8\linewidth]{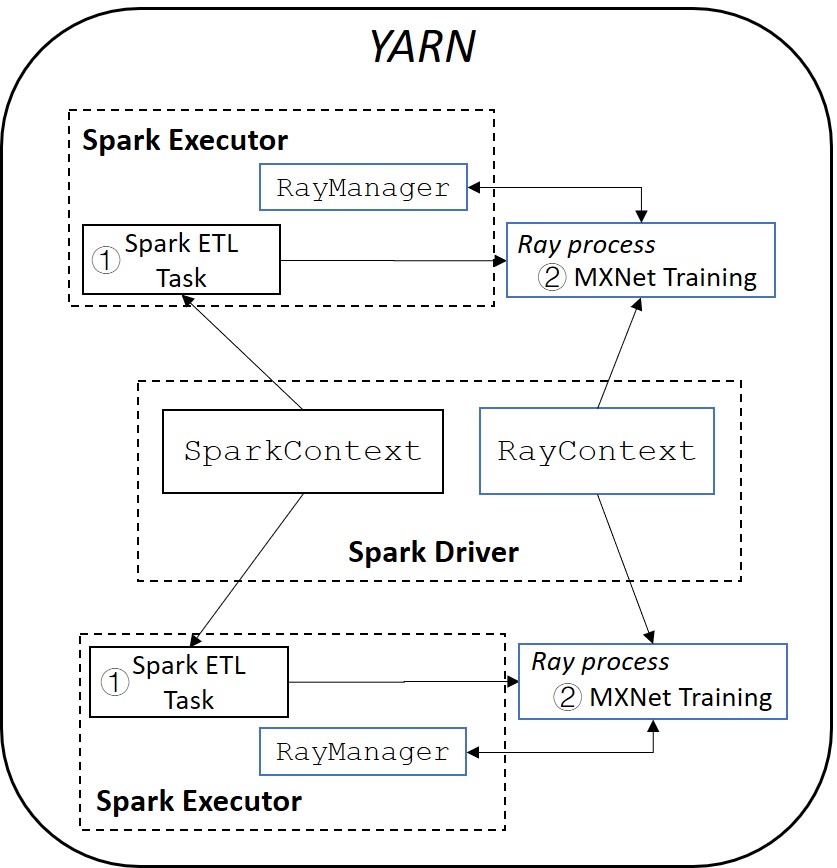}
\caption{RayOnSpark architecture overview.}
\label{fig:rayonspark}
\end{figure}

On a shared Hadoop YARN \cite{vavilapalli2013apache} cluster, we first use Spark \cite{zaharia2012resilient} to process the data; then we launch a Ray \cite{moritz2018ray} cluster alongside Spark using RayOnSpark to run distributed deep learning training directly on top of the YARN cluster. 

We design \emph{RayOnSpark} for users to run distributed Ray applications on big data clusters. Figure \ref{fig:rayonspark} illustrates the architecture of \emph{RayOnSpark}. In the Spark implementation, a Spark program runs on the driver node and creates a \emph{SparkContext} object, which is responsible for launching multiple Spark executors on the YARN cluster to run Spark jobs. In \emph{RayOnSpark}, the Spark driver program can also create a \emph{RayContext} object, which will automatically launch Ray processes alongside each Spark executor; it will also create a \emph{RayManager} inside each Spark executor to manage Ray processes (e.g., automatically shutting down the processes when the program exits). As a result, one can directly write Ray code inside the Spark program. The processed Spark’s in-memory DataFrames or Resilient Distributed Datasets (RDD) \cite{zaharia2012resilient} can be directly fed into the Ray cluster through the Plasma object store used by Ray for distributed training.

We choose Apache MXNet \cite{chen2015mxnet} as our deep learning framework. On top of \emph{RayOnSpark}, we implement a lightweight shim layer to automatically deploy distributed MXNet training on the underlying YARN cluster. Similar to RaySGD \cite{RaySGD}, each MXNet worker (or parameter server) runs as a Ray actor \cite{moritz2018ray}, and they directly communicate with each other via the distributed key-value store provided natively by MXNet. In this way, users are relieved from managing the complex steps of distributed training, through the scikit-learn \cite{pedregosa2011scikit} style APIs provided by the shim layer. In addition, the shim layer also provides just-in-time Python package deployment across the cluster, so that the user no longer needs to install the runtime environment for each application on each node beforehand, and the cluster environment remains clean after the job finishes. The same methodology can be easily applied to other deep learning frameworks such as TensorFlow \cite{abadi2016tensorflow} and PyTorch \cite{paszke2017automatic}.

\subsection{Model Training Pipeline.}

With \emph{RayonSpark}, we now simplify our end-to-end model training process by combining Spark ETL job and distributed MXNet training job into one unified pipeline. Our model training process starts from launching Spark tasks to extract our restaurant transactions data stored on distributed file systems. We then perform data cleaning and preprocessing steps on the YARN cluster. Once the Spark job has completed, the Spark driver program launches Ray on the same YARN cluster and dynamically distributes MXNet packages across the cluster. \emph{RayonSpark} helps to set up the distributed deep learning environment and launches MXNet processes alongside Spark executors; each every MXNet worker takes the partition of the preprocessed data from Ray’s Plasma object store on its local node and starts model training. The trained model is then saved with a version tag for online serving.

Listing \ref{code} shows the code segments of our distributed MXNet training pipeline. As described in Section \ref{rayonspark}, we first initiate the \emph{SparkContext} on YARN and use Spark to do data processing followed by initiating the \emph{RayContext} to launch Ray on the same cluster. We implement a scikit-learn \cite{pedregosa2011scikit} style \emph{Estimator} for MXNet which takes the MXNet model, loss, metrics and training config (e.g., the number of workers and parameter servers) as inputs. MXNet \emph{Estimator} performs the distributed MXNet training on Ray by directly fitting Spark DataFrames or RDDs \cite{zaharia2012resilient} to the given model and loss. In our implementation, only several extra lines of code are needed to scale the training pipeline from a single node to production clusters.

\begin{lstlisting}[label=code, language=Python, caption=Code sample for data  processing and training pipeline with \emph{RayOnSpark}.]
sc = init_spark_on_yarn(...)
# Use SparkContext to load data as Spark Dataframe or RDD, and do data processing.

ray_ctx = RayContext(sc)
ray_ctx.init()

mxnet_estimator = Estimator(model, loss, metrics, config)
mxnet_estimator.fit(train_rdd, val_rdd, epochs, batch_size)
\end{lstlisting}

This unified software architecture allows big data processing and deep learning training workloads to run on the same big data cluster. Consequently, we only need to maintain one big data cluster for the entire AI pipeline, with no extra data transfer across different clusters. This achieves the full utilization of the cluster resources and significantly improves the end-to-end performance of the whole system.

\section{Experiment Results}
In this section, we experimentally evaluate the T\emph{x}T model through both offline evaluation and online A/B testing in Burger King’s production environment.

\subsection{Setup.}

\emph{Dataset}. The dataset is constructed from Burger King’s customer drive-thru transaction records of the past year. The historical data of the first 11 months is used as training data and the last month is for validation. An add-to-cart event sequence is formed for each order from the transactions and the last item in the sequence is used for prediction. We select weather descriptions, order time, current temperature and a number of restaurant attributes including identifier and location as context features.

\emph{Baselines}. To evaluate the effectiveness of the T\emph{x}T model, we select several of its variations as baseline models for comparison, i.e. RNN \cite{Hidasi2016SessionbasedRW}, RNN Latent Cross \cite{beutel2018latent}, and a variation of ItemCF model \cite{sarwar2001item} with multiplicative to context features (Contextual ItemCF). Table \ref{tab:model} summarizes whether these models consider order sequence and context features.

\begin{table}
  \caption{Model variations for comparison.}
  \label{tab:model}
  \begin{tabular}{|p{3cm}|c|c|}
    \hline
    \textbf{Model} & Sequence Input & Context Inputs\\
    \hline
    RNN & \checkmark & \\
    \hline
    RNN Latent Cross & \checkmark & \checkmark\\
    \hline
    Contextual ItemCF & & \checkmark\\
    \hline
    T\emph{x}T & \checkmark & \checkmark\\
    \hline
  \end{tabular}
\end{table}

\emph{Evaluation metrics}. For offline measurement, we use Top1 and Top3 accuracy to evaluate the performance of different models since for drive-thru, three food recommendations are shown simultaneously to the guest on the digital menu board with the first recommendation highlighted. For online measurement, we conducted online A/B testing in our production environment and use conversion gain (observed recommendation conversion uplift over the control group) and add-on sales gain (observed add-on sales uplift over the control group) to evaluate the model performance.

\begin{figure}[ht]
\centering
\includegraphics[width=0.9\linewidth]{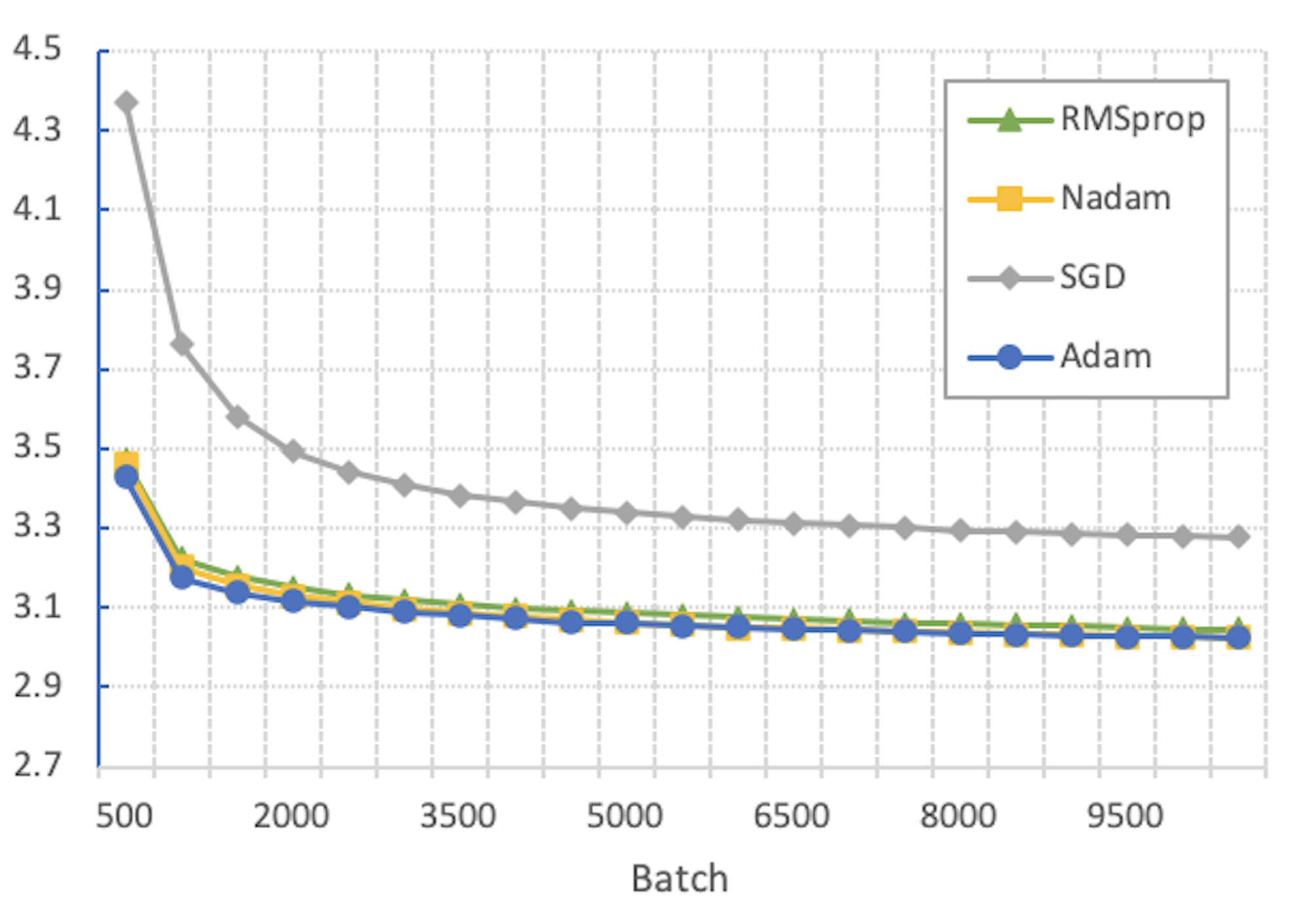}
\caption{Training loss of T\emph{x}T with different optimizers.}
\label{fig:optimizer}
\end{figure}

\emph{Environment and Configurations}. 
Our model is implemented with Python 3.6 and MKL-DNN optimized version of Apache MXNet (i.e. mxnet-mkl) 1.6. Based on our experiential results shown in Figure \ref{fig:optimizer}, we choose Adam \cite{kingma2014adam} as the optimizer for all models. The training configurations of T\emph{x}T are shown in Table \ref{tab:config}. Other baseline models use the compatible set of the corresponding configurations as T\emph{x}T.

\begin{table}
  \caption{Training configurations of T\emph{x}T.}
  \label{tab:config}
  \hskip-0.5cm\begin{tabular}{|p{2.9cm}|p{0.62cm}|p{2.9cm}|p{0.62cm}|}
  \hline
   \multicolumn{2}{|c}{\emph{Sequence Transformer}} &
   \multicolumn{2}{|c|}{\emph{Context Transformer}} \\
  \hline
  Embedding size & 100 & Embedding size & 100 \\
  \hline
  Head number & 4 & Head number & 2 \\
  \hline
  With Mask & True & With Mask & False \\
  \hline
  Transformer layers & 1 & Transformer layers & 1 \\
  \hline
  Sequence length & 5 & \multicolumn{2}{c|}{} \\
  \hline
   \multicolumn{2}{|l}{Batch size} &
   \multicolumn{2}{|l|}{512} \\
  \hline
   \multicolumn{2}{|l}{Epochs} &
   \multicolumn{2}{|l|}{1} \\
  \hline
   \multicolumn{2}{|l}{Learning rate} &
   \multicolumn{2}{|l|}{0.001} \\
  \hline
  \end{tabular}
\end{table}

\subsection{Comparative Analysis.}
We conducted offline comparison for model training using the above evaluation metrics and configurations. We also conducted online A/B testing of different recommendation systems for 8 weeks. We share our results in this subsection.

\begin{figure}[ht]
\centering
\includegraphics[width=0.9\linewidth]{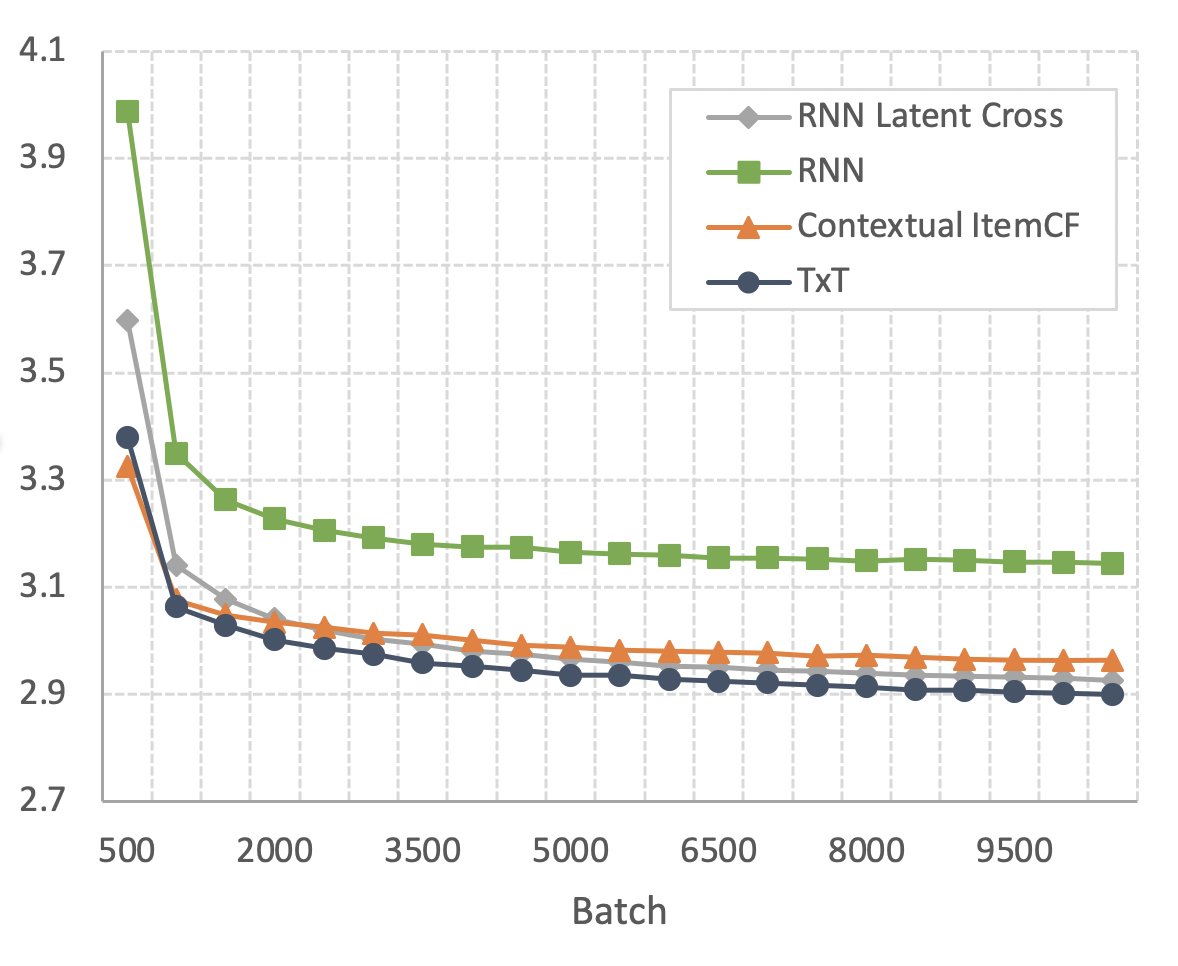}
\caption{Training loss of model variations.}
\label{fig:loss-model}
\end{figure}

As shown in Figure \ref{fig:loss-model} and Table \ref{tab:offline}, when only sequence information (using RNN) or context information (using Contextual ItemCF) is considered, we observe significantly higher cross entropy loss during training and lower validation accuracy compared with models that take both into consideration (using RNN Latent Cross and T\emph{x}T). Compared with RNN Latent Cross, T\emph{x}T is able to further improve the Top1 and Top3 accuracy by around 1.4\% and 2.4\% respectively.

\begin{table}
  \caption{Comparison of offline training results.}
  \label{tab:offline}
  \hskip-0.8cm\begin{tabular}{p{3.5cm} r r}
    \hline
    \textbf{Model} & Top1 Accuracy & Top3 Accuracy\\
    \hline
    RNN & 29.98\% & 46.24\%\\
    \hline
    Contextual ItemCF & 32.18\% & 48.37\% \\
    \hline
    RNN Latent Cross & 33.10\% & 49.98\% \\
    \hline
    T\emph{x}T & 34.52\% & 52.37\%\\
    \hline
  \end{tabular}
\end{table}

For online measurement, due to time constraints and business requirements, we divided online comparison into two different phases. In the first stage, we measured T\emph{x}T against an existing blackbox recommendation system provided by an outside vendor that we have been previously using. We randomly divided our pilot restaurants into two groups and ran both systems simultaneously for 8 weeks. As shown in Table \ref{tab:online}, T\emph{x}T model improved the recommendation conversion rate for drive-thru by +79\%. Production observation also shows that T\emph{x}T is able to react to location and weather changes very well. In the second phase, we conducted online A/B testing for model variations at our drive-thru environment for 4 weeks. For the control group, 50\% of the drive-thru transactions were randomly selected to use the RNN Latent Cross model while the remaining transactions used the T\emph{x}T model. It turned out that T\emph{x}T was able to improve the conversion rate by 7.5\% with 4.7\% add-on sales gain over the control group.

\begin{table}
  \caption{Comparison of online A/B testing results.}
  \label{tab:online}
  \begin{tabular}{|p{3cm}|p{2cm}|p{2cm}|}
  \hline
  Model & Conversation Rate Gain & Add-on Sales Gain  \\
  \hline
   \multicolumn{3}{|l|}{\textbf{\emph{Phase 1}}} \\
  
  \hline
  Blackbox vendor's system (control) & - & - \\
  \hline
  T\emph{x}T & +79\% & +14\%\\
  \hline
   \multicolumn{3}{|l|}{\textbf{\emph{Phase 2}}} \\
  \hline
  RNN Latent Cross (control) & - & - \\
  \hline
  T\emph{x}T & +7.5\% & +4.7\%\\
  \hline
   \multicolumn{3}{|l|}{\textbf{\emph{Additional A/B Testing on Mobile APP}}} \\
   \hline
  Rule-based system (control) & - & - \\
  \hline
  {\small Google Recommendation  AI} & +164\% & +64\% \\
  \hline
  T\emph{x}T & +264\% & +137\%\\
  \hline
  \end{tabular}
\end{table}

To further test T\emph{x}T’s performance, we also ran our recommendation system on Burger King’s mobile application for 4 weeks side by side with Google Recommendation AI, a state-of-art recommendation service provided by Google Cloud Platform. For the control group, we randomly selected 20\% of the users and presented them with a previous recommendation system based on simple rules. As shown in Table \ref{tab:online}, T\emph{x}T improved recommendation conversion at the checkout page by 264\% and add-on sales by 137\% when compared with the control group. This also stands for +100\% conversion gain and +73\% add-on sales gain when compared with the test groups running Google Recommendation AI service.

\section{Experience}
We have deployed the end-to-end context-aware recommendation system in Burger King’s selected drive-thru restaurants for 10 months, and also deployed our T\emph{x}T model on Burger King’s mobile application for product recommendation at check-out. In this section, we share some key learnings and insights we have gained when building and deploying our recommendation system.

In drive-thru or other offline customer interaction channels where limited user data could be obtained, guest order sequence and different context features can be exploited for recommendation. Transformer encoder and mean-max pooling architectures can well extract the feature representations of these inputs. Using our T\emph{x}T model, we have seen 79\% improvement in real-world conversion rate for drive-thru compared with the existing vendor’s recommendation solution. In addition, our T\emph{x}T model has also proven to be effective even applied to online settings. The preliminary results show that T\emph{x}T outperforms other test groups using out-of-box solutions such as Google Recommendations AI on both recommendation conversion rate and add-on sales.

In the real production environment, building a unified end-to-end system for the entire pipeline is critical to save time, cost and efforts. Previously we used a separate GPU cluster for model training and nearly 20\% of the total time is spent on data transfer to the GPU cluster. After adopting RayOnSpark described in Section \ref{rayonspark}, we find it far more productive to combine big data processing and deep learning tasks in a single end-to-end pipeline. Using the implementation with high-level APIs in project \cite{AnalyticsZoo}, engineers already with Spark experience can quickly learn how to inject AI workloads to the existing clusters in a distributed fashion.

\begin{figure}[ht]
\centering
\includegraphics[width=1.0\linewidth]{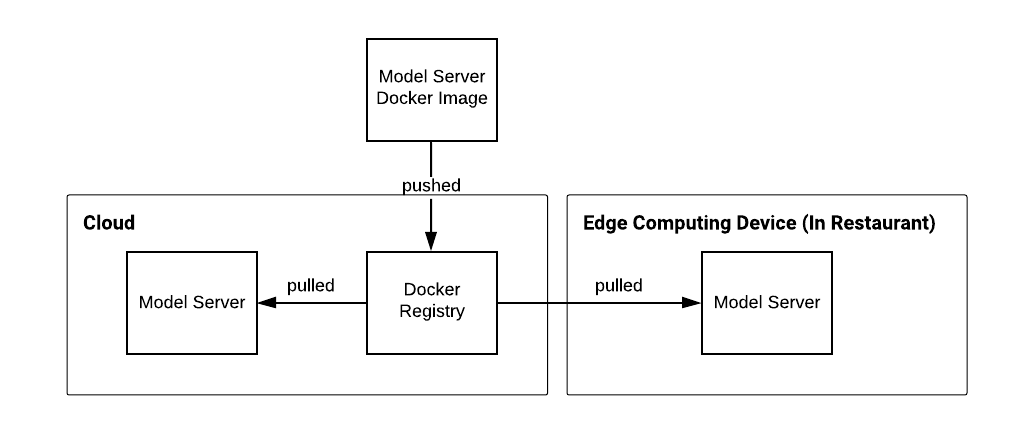}
\caption{Model Serving Architecture.}
\label{fig:serving}
\end{figure}

Our drive-thru recommendation system needs to be able to provide recommendations in sub milliseconds to all of Burger King restaurants across different locations. As network conditions vary from restaurant to restaurant, we design our system in a way such that it can either be deployed to cloud or on-premise to ensure high availability to our drive-thru guests. Figure \ref{fig:serving} shows how we manage our models using Docker registry; both cloud and on-premise systems can pull latest model images from the docker registry to run recommendation inference. This setup combined with T\emph{x}T model’s location awareness eliminates the need to manage different models across multiple restaurant regions. Self-contained model servers can also be deployed to any environment with Docker installed.

\section{Conclusion}
In this paper, we introduced the context-aware drive-thru recommendation system implemented at Burger King. Our \emph{Transformer Cross Transformer} model (T\emph{x}T) has been proven to be effective in the drive-thru circumstance for modeling user order behavior and complex context features. Experimental results show that T\emph{x}T outperforms other existing recommendation solutions in Burger King’s production environment. We further implemented a unified system based on Spark and Ray to easily run data processing and deep learning training tasks on the same big data cluster. Our solution has been open sourced in \cite{AnalyticsZoo} and we shared our practical experience for building such an end-to-end system, which could be easily applied to other customer interaction channels.

\bibliographystyle{siamplain}
\bibliography{reference}

\end{document}